\documentclass[twocolumn,showpacs,preprintnumbers,amsmath,amssymb]{revtex4}

\usepackage{epsfig}
\usepackage{graphicx}
\usepackage{dcolumn}
\usepackage{bm}
\usepackage{amsmath}
\usepackage{float}

\include{math}
\include{theorems}

\begin{document}


\title{Limited profit in predictable stock markets}

\author{Roland Rothenstein}
\author{Klaus Pawelzik}%

\affiliation{%
Institute for Theoretical Physics, \
University of Bremen,  FB 1 \\
Otto-Hahn-Allee, 28334 Bremen \\
}%

\date{\today}

\begin{abstract}
It has been assumed that arbitrage profits are not possible in
efficient markets, because future prices are not predictable. Here
we show that predictability alone is not a sufficient measure of
market efficiency. We instead propose to measure inefficiencies of
markets in terms of the maximal profit an ideal trader can take
out from a market. In a stock market model with an evolutionary
selection of agents this method reveals that the mean relative
amount of realizable profits $P$ is very limited and we find that
it decays with rising number of agents in the markets.
Our results show that markets may self-organize their collective
dynamics such that it becomes very sensitive to profit attacks
which demonstrates that a high degree of market efficiency can
coexist with predictability.
\end{abstract}


\maketitle

Many complex systems in nature and society consist of rather
simple elements whose interactions self-organize in a way that the
available dynamics of the system achieves a desired performance.
In brains, for instance, individual neurons can realize only
rather simple functions while their interactions are permanently
re-adapted to ultimately yield successful behavior in
non-stationary environments. Further examples are found in the
metabolic and genetic networks in cells \cite{Segre_02}. It has
been suspected, that also collective human behavior can sometimes
be described in a similar way \cite{Barabasi_02}. A particularly
prominent example for social interactions is a market in which
each trader has only limited access to relevant information, while
the dynamics of the whole system is considered to exploit the
overall information quite effectively \cite{Chen_03}. In economy
it is believed that the collective behavior of traders leads to an
efficient market such that in real markets no profit can be made
from an analysis of price time series \cite{Samuelson_65,
Fama_70}. However, since high frequency data are available
research has revealed that some temporal correlations in price
time series do in fact exist. But it is still an open question
whether these correlations imply inefficiencies (see \cite{Lo_ME}
for an overview). For a solution of this problem it would be
highly desirable to have a quantitative measure of inefficiency
\cite{Farmer_99}, for which, however, no generally accepted
definition has been proposed until now.

Former studies of efficiency were challenged by possible
irrationalities of traders in the market. They either postulated a
market of rational traders or tried to measure efficiency of real
markets \cite{Lo_ME}. The drawback of these approaches is that in
the first a comparison to real markets in principle cannot be
complete, while the second lacked a clear definition of irrelevant
noise and meaningful information.

In this paper we propose a novel measure of  inefficiency and
apply it to a model which combines the approaches of evolutionary
games \cite{Hofbauer_Games} and agent-based models \cite{Levy}.
The method we propose for measuring the inefficiency of a
dynamical system modeling a market is general and can be applied
to other models and in principle also to real markets. In our view
our approach is the first step towards an objective measure of
efficiency of stock markets.

Our starting point is to consider a market as an open dynamical
system $\mu$.
The future return $r_{t+1}= (p_{t+1}- p_t)/p_t$
then is given by
\begin{equation}
r_{t+1}=\mu(\bf{\Phi_t}),
\end{equation}
where $p_t$ is the price at time $t$. $\bf{\Phi_t}$ denotes all
parameters determining the next price including the actions of the
traders participating in the market at time $t$.

This formulation of a market shows why it is not straightforward
to use the notion of predictability for defining efficiency: a
dynamical system can in principle be perfectly predicted if all of
its constituents are known.
Making profit, however, for every trader requires to interact with
the dynamics of a market. If an external additional trader enters the
market at time $t$ her interaction $\bf W$ will influence the
resulting returns
\begin{equation}
\label{Eqn:newreturn} \tilde r_{t+1} = \mu'(\bf{\Phi}_{t},\bf W)
\end{equation}
and will thereby also modify the profit the trader can achieve.
Note that $\mu'(\bf{\Phi}_{t},\bf 0) = \mu(\bf{\Phi}_{t})$.

For the purpose of analyzing a given market model we can formally
close a given system with respect to capital and information by
using a procedure analogous to the notion of baths in
thermodynamics. This is done by assuming that the system is
coupled to an ideal trader \textbf{T} who has perfect knowledge of
the system and infinite resources. Furthermore, we assume
\textbf{T} always chooses her interaction with the market such
that she achieves the maximal profit possible. Clearly, the ideal
situation of complete information is very artificial and in
reality only a small part of the information determining the
dynamics $\mu$ is available to a trader. However, the profit of
this ideal trader per definition provides a strict upper bound on
the profit achievable by any trader.

In a simple stock market with one risk less asset M e.g. cash and
a single stock S the optimal interaction of the ideal trader can
be represented by the amount of $M$ and a price limit $p$ for
buying or selling, i.e. ${\bf W} = (M,p)$. In the following we
assume that \textbf{T} always chooses the optimal value for $p$
such that the strength of interaction is characterized solely by
the magnitude of risk less assets $M$ she is willing to exchange.

The ideal trader \textbf{T} would buy stocks with $M>0$, if the
return $\tilde r_{t+1}$ resulting from her interaction would be
positive. If in contrast the return $\tilde r_{t+1}$ resulting from
an interaction would be negative the ideal trader sell her stocks with
$M<0$. The profit is calculated according to the difference between
buying and selling at time $t+1$:

\begin{equation}
\label{Eqn:Profit}
P_t(M)=\frac{M}{M_{tot}}\tilde{r}_{t+1}(M).
\end{equation}
$M=p S$ denotes the value of the stocks at time $t$ the
trader buys or sells measured in risk less assets. $M_{tot}$
denotes the total value of risk less assets present in the market.
The profit is equivalent to the amount of risk less assets
$\frac{\Delta M}{M_{tot}} = \frac{M_{t+1}-M_t} {M_{tot}}$ the
ideal trader could virtually win in one time step. This can be
seen as the relative loss of money of the participants in the
market potentially transferred to the ideal trader.

Using all information, the ideal trader will also take into
account her own influence on the market \cite{Challet_00} in order
to optimize her profit $P_t$ in every time step $t$:
\begin{equation}
\label{Eqn:Maxi} \hat{P_t} = P_t(\hat{M_t})=\max_{M} P_t(M).
\end{equation}

Assuming that $P_t(M)$ is differentiable we obtain for the optimal
invested capital
\begin{equation}
\label{Eqn:maxM}
\hat{M_t}=-\frac{\tilde{r}_{t+1}(\hat{M_t})}
                {\partial_M \tilde{r}_{t+1}(\hat{M_t})}.
\end{equation}
For the optimal profit then follows:
\begin{equation}
\label{Eqn:maxProfit}
\hat{P}_t(\hat{M_t})=-\frac{\tilde{r}_{t+1}^2(\hat{M_t})}
{\partial_M  \tilde{r}_{t+1}(\hat{M_t})}
\end{equation}
This equation shows that the optimal profit consists of two
contributions: The nominator represents the volatility of the
return, resulting from the interaction with \textbf{T}. The
denominator characterizes the responsiveness of the market to
increasing $M$, and reflects the sensitivity of the system.

\begin{figure}[ht]
\begin{center}
\includegraphics[width=6cm,angle=270]{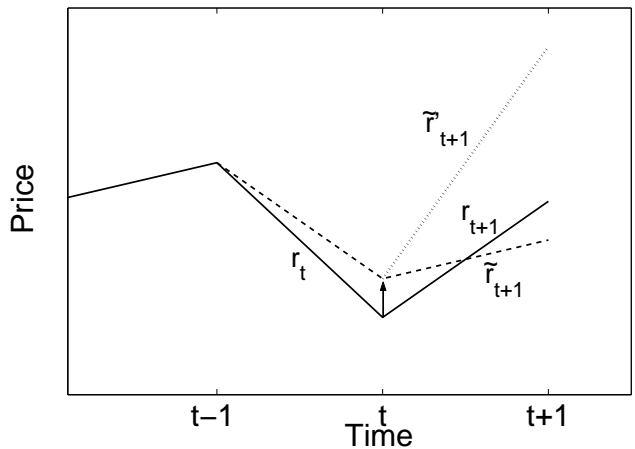} \hfill
\caption{\small Example of a price time series (solid line) and
possible changes of it, if an ideal trader buys stocks at time
$t$. When an additional buy order is placed, the price at time $t$
will always be higher
than without interaction (arrow at $t$).
The value of the returns at $t+1$ depends on the market dynamics
and reflects the rationality of the traders. The dashed line shows
an example characteristic for a defensive market in which
$\delta_t(M) <0$. The dotted line shows a price history for a
non-defensive reaction ($\delta_t(M) >0$).} \label{Abb:Return}
\end{center}
\end{figure}

The main problem of determining the inefficiency of a market is to
specify $\tilde r(M)$. One can in principle estimate this function
in experiments or measure it in model markets (see below) or one
can postulate generic properties of this function and analyze the
effect on the inefficiency \cite{Pawelzik_??}. We define the
defensiveness of a market as $ \delta_t(M)=\frac{r_{t+1}}{|r_{t+1}|}
(\tilde r_{t+1}(M)-r_{t+1})$. One might expect that for an optimal order
$|\hat M| > 0$, the market reaction will be defensive and $\delta$
will be negative (see in Fig. \ref{Abb:Return} the dashed line).
While for large orders any realistic market, as e.g. the example
below, will ensure this, for intermediate orders and irrational
traders a market can be non-defensive ($\delta>0$)(see in Fig.
\ref{Abb:Return} the dotted line). In the latter case, the ideal
trader will maximally exploit these irrationalities of a market
which leads to the counterintuitive result that a feedback
of knowledge into a market can in principle increase its overall volatility.\\

We illustrate the determination of $\tilde r$  and how one can
measure the efficiency using a simple order-based stock market
model. Our model reproduces the statistics of empirical market
returns to an astonishing large degree, including characteristic
scaling of return distributions and strong temporal correlations
of volatility (volatility clusters). Because this model and it's
properties have been described in detail before
\cite{Rothenstein_2002_PhysicaA} we here outline only its main
constituents.

In our model $N$ agents trade by swapping stocks into cash and
vice versa. Initially every agent $i$ receives $S_0$ stocks and an
amount $M_0$ of cash. The decision of each agent to buy or to sell
stocks is determined by a set of $n$ parameters $\alpha_{\Delta
t}^i (\Delta t=0,...,(n-1))$ and is based on the history of the
returns $r_{t-1}, r_{t-2}, ..., r_{t-n+1}$. The parameters define
simple linear prediction models for each agent and are randomly
drawn from a normal distribution with mean 0 and variance 1.

Every iteration of the model yields the next return. It can be
divided in three parts: First, the agents make a prognosis of two
successive future returns $\hat r_t, \hat r_{t+1}$ based on their
individual parameters of the linear prediction model.
\begin{eqnarray*}
    \hat{r}^i_t &=& \alpha^i_0
        +\sum^{\tau}_{\Delta t=1} \alpha^i_{\Delta t} r(t-\Delta t)\\
    \hat{r}^i_{t+1}&=& \alpha^i_0 + \alpha^i_1 \hat{r}^i_t
        + \sum^{\tau}_{\Delta t=2} \alpha^i_{\Delta t} r(t-\Delta t+1).
\end{eqnarray*}

Then the agents decide to buy/sell stocks if they predict that the
second return $\hat r_{t+1}$ is positive/negative. In the second
part each agent $i$ places an order, which is stored in an order
book. The overall demand $D(p)$ and supply $O(p)$ of the stocks is
calculated according to a limit price given by the agents and the
size of the orders. $S^i$ is the total number of stocks agent $i$
place in its order respectively $\Delta S^i=$
int$[\frac{M^i}{\hat{p}^i_t}]$ the number of stocks agent $i$ is
willing to buy with her money $M^i$:
\begin{eqnarray*}
O(p)&=&\sum_{i: \hat{r}^i_{t+1} <0} S^i \Theta(p-\hat{p}^i_t) \\
D(p)&=&\sum_{i:\hat{r}^i_{t+1} > 0} \Delta S^i \Theta(\hat{p}^i_t-p),
\end{eqnarray*}
where $\Theta(x)$ is the Heaviside step function. After the orders
are placed, a turnover function is calculated. That is the minimum
of both functions at price $p$:
\begin{eqnarray*}
Z(p)=\min\{O(p),D(p)\}
\end{eqnarray*}

In order to determine the new price the minimum and the maximum
argument of $Z(p)$ at the interval of the maximum turnover
$p_{\min} = \min( $argmax $ Z(p))$ and $p_{\max} = \max( $argmax
$Z(p))$ are computed and the new price is then defined by the
weighted mean between these two points:
\begin{eqnarray*}
p(t)=\frac{p_{\min}O(p_{\min})+p_{\max}D(p_{\max})}{O(p_{\min})+
D(p_{\max})}
\end{eqnarray*}

\begin{figure}[ht]
\begin{center}
\includegraphics[width=6cm,angle=270]{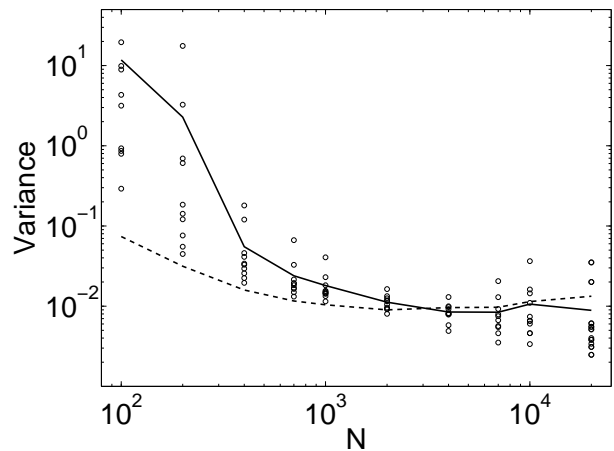} \hfill
\caption{\small Variance $\langle r_t^2 \rangle$ of the returns in
the undisturbed model versus number of agents $N$ (dashed line).
The circles indicate the variances $\langle \tilde r_t^2 \rangle$
of 10 runs of the system interacting with the ideal trader
\textbf{T} averaged from 500 simulation steps. Full line: variance
averaged from the 10 simulations (circles).} \label{Abb:VarvsN}
\end{center}
\end{figure}

In the last step of each iteration one agent is replaced by a new
random one. In case of an evolutionary update the poorest agent is
replaced. To investigate the effects of evolution on efficiency we
also consider the case where a randomly chosen agent becomes
replaced in each step. Previously we have shown that for large $N$
the return distributions have a stationary and characteristic
shape. While the shape of the distributions varies for the
different evolutionary mechanisms
\cite{Rothenstein_2002_PhysicaA}, the variances $\langle r_t^2
\rangle$ for both cases are nearly equal and enter a constant
value for sufficiently many traders $N
> 1000$ (see dashed line in Fig. \ref{Abb:VarvsN}).

The ideal trader applied to our model maximizes the amount of
money she virtually gets from the system without giving stocks
away. The only free variable the ideal trader can optimize is the
money she uses to buy or sell shares. Following Eqn.
\ref{Eqn:Profit} she maximizes her profit $P_t(M_t)$. To do this,
the trader has to place an order in the order book. For every
price $p$ one can calculate a spread $\sigma(p)$ between offered
shares $O(p)$ and demanded shares $D(p)$:
\begin{equation*}
\sigma(p)= O(p) - D(p)
\end{equation*}
If at time $t$ \textbf{T} would buy $S=$int$[\frac{M}{p}]$ shares,
she placed an order at the lowest $p$ with $S \ge \sigma(p)$. This
leads to a new demand function and therefore to a new price
$\tilde p_{t+1}$. The procedure for selling is analogous. To
calculate the optimal profit $\hat P$ (see Eqn. \ref{Eqn:Maxi})
the ideal trader repeats this procedure for all possible values of
$M$, and finally invests the optimal amount of capital $\hat M_t$.
Per construction \textbf{T} does not further interact with the
market in the subsequent time step, which makes the profit
$\hat{P_t}$ in fact virtual. After estimating $\hat P_t$ the
influence of \textbf{T} on market is canceled and the method is
applied de novo in the next time step. For our market model this
approach corresponds to an interaction which does not change the
nature of the system. It provides a strict upper bound on the
profit achievable by any realistic trader who in contrast would
need to interact with the market a least twice (e.g first buy and
then sell). Our method avoids this more realistic scenario because
such an interaction would effectively remove stocks (or money)
from the market which can change the dynamics of the system
substantially and thereby will prevent a reliable characterization
of the market's properties.

\begin{figure}[ht]
\begin{center}
\includegraphics[width=6cm,angle=270]{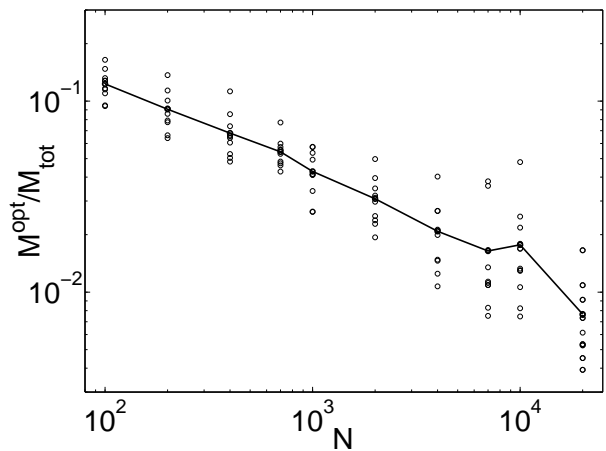} \hfill
\hfill \caption{\small The mean relative size of an optimal order
$\langle \frac{\hat M}{M_{tot}} \rangle$\ (measured in risk less
assets) versus the number of traders $N$. Circles mark the mean
size of one simulation run calculated from 500 simulation time
steps. The line is the mean from the different simulation runs.}
\label{Abb:MaxKvsN}
\end{center}
\end{figure}

\begin{figure}[ht]
\begin{center}
\includegraphics[width=6cm,angle=270]{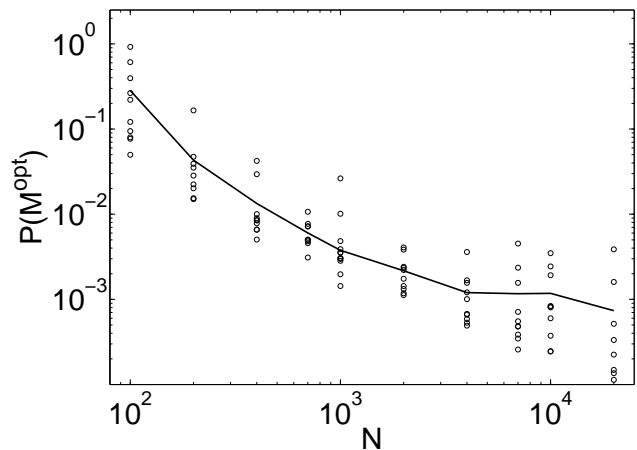} \hfill
\caption{\small Mean of the optimal profit $\hat P$ an ideal
trader can achieve in the market depending on the number of
agents. Circles show the mean optimal profit averaged over 500
simulation time steps. The line is the mean over this 10
simulations.} \label{Abb:PvsN}
\end{center}
\end{figure}

In Fig. \ref{Abb:VarvsN} one can see that especially in the case
of small $N$ the volatility becomes strongly amplified, while for
larger N the market becomes much more defensive. Interestingly,
there is no significant difference in the amplification of the
cases with and without evolution (not shown).

Fig. \ref{Abb:MaxKvsN} shows that the mean of the optimal
investment in units of the total money in the market drops with a
power law, i.e. $ \frac{|\hat M_t|}{M_{tot}}\propto N^{-1/2}$. We
find that the ideal invested capital in a market without evolution
always lies above the market with evolution and on average is a
factor of $1.39 \pm 0.35$ greater (not shown). Both results
clearly have an influence on the mean profit $P$ an ideal trader
can achieve in our market (see Fig. \ref{Abb:PvsN}). We find that
the possible profit $P$ in a market with evolution is for all $N$
smaller and is on average only $44\%\pm29\%$ of the average profit
$P$ of the case of random agent replacement. Eqn.
\ref{Eqn:maxProfit} indicates that this effect is mainly caused by
the sensitivity $\tilde r'(M)$ of the system because the variances
do not differ in both markets. Furthermore, we see that the
highest achievable profit is decreasing for a rising number of
agents. In the regime $N>1000$ the decrease of P is mainly due to
the fact that $\hat M$ is getting smaller. Most striking are the
absolute values of P. Taking into account that the agents of our
market model are not very smart, the fraction of capital the ideal
trader can get in one time step on average is below 1\% in the
case of 1000 agents and below 0.1\% for 20000 agents. Our results
furthermore indicate that the profit \textbf{T} can achieve in the
market might fall with a power law $\propto N^{-1/2}$ which would
imply a vanishing profit opportunity for any trader in the
thermodynamic limit.

Prediction of financial time series is an interesting and
challenging topic of research on its own. Our analysis, however,
shows that predictability alone is not sufficient to characterize
the efficiency of a market. Markets can have different degrees of
defensiveness and sensitivity, which limits the profit possible
for any trader. Our model not only illustrates the notion of
the ideal trader, but shows that the profit option for a trader
might vanish if the market is large. Furthermore, the comparison
of the evolutionary market with the case of random replacements of
agents underlines the long standing hypothesis that efficiency of
markets might emerge from Darwinian mechanisms. Finally, our
measure of inefficiency is general enough to be applicable to
every well defined market model and can be used to analyze simple
models \cite{Pawelzik_??}. We believe our novel method to measure
market inefficiencies paves the way towards a systematic
understanding of the influences of irrationalities of traders and
we expect it will help to better understand the dynamics of real
markets.

\bibliographystyle{h-physrev}
\bibliography{ownpub,finance}

\end{document}